# Enabling plastic co-deformation of disparate phases in a laser rapid solidified Sr-modified Al-Si eutectic through partial-dislocation-mediated-plasticity in Si

Arkajit Ghosh[a*], Wenqian Wu[b], Bibhu Prasad Sahu[a], Jian Wang[b], Amit Misra[a,c*]

[a] Department of Materials Science and Engineering, University of Michigan – Ann Arbor, MI 48109, USA
[b] Department of Mechanical and Materials Engineering, University of Nebraska – Lincoln, NE 68588, USA
[c] Department of Mechanical Engineering, University of Michigan – Ann Arbor, MI 48109, USA

*Corresponding authors' email addresses: arkajitg@umich.edu; amitmis@umich.edu

## Abstract

Nano-scale eutectics, such as rapid solidified Al-Si, exhibit enhanced yield strength and strain hardening but plasticity is limited by cracking of the hard phase (Si). Mechanisms that may suppress cracking and enable plastic co-deformation of soft and hard phases are key to maximizing plasticity in these high-strength microstructures. Using a combination of laser rapid solidification and chemical (Sr) modification, we have synthesized fully eutectic Al-Si microstructures with heavily twinned Si nano-fibers that exhibit high hardness up to 2.9 GPa, and high compressive flow strength (~840 MPa) with stable plastic flow to ~26% plastic strain. After deformation, the hard Si(Sr) fibers did not exhibit cracks, but a high density of stacking faults were observed in the Si(Sr) fibers suggesting partial dislocation mediated plasticity. Mechanisms for suppression of cracking and activation of partial dislocations in Si deformed at room temperature are discussed in terms of nanoscale fiber geometry with reduced aspect ratio and lowering of the Peierls barrier in chemically-modified, nano-twinned Si fibers.

## Highlights

- Ultrafine Al-Si eutectic with heavily nano-twinned Si fibers have been synthesized by fine-spot laser surface remelting of 0.2 wt% Sr modified Al - 20 wt% Si alloy.

- The ultrafine microstructure exhibits extraordinary combination of micromechanical properties: up to 2.9 GPa of nanoindentation hardness and ~840 MPa of compressive strength with a stable plastic flow of up to 26%.

- Disparate eutectic phases, i.e., soft Al and hard Si, plastically co-deforms. Deformation mechanisms in Al and Si have been established as confined channel slip (CCS) and partial dislocation mediated plasticity (PDMP).

- Huge strain hardening comes from shearing of Al nano-channels (CCS), while PDMP in Si suppresses fiber fracture and thereby microstructural failure until a large plastic strain.

# Graphical abstract

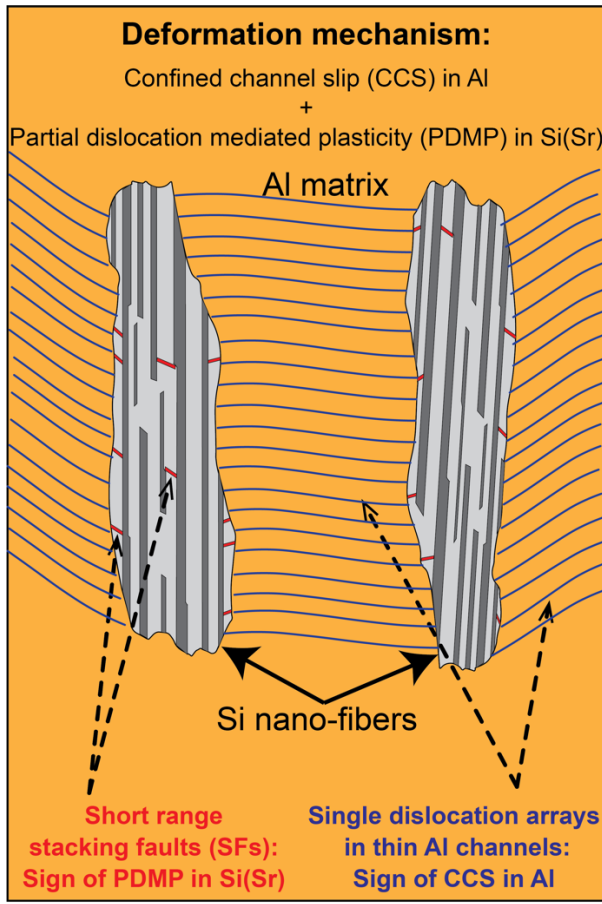
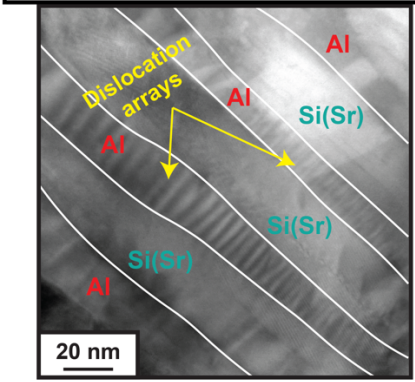
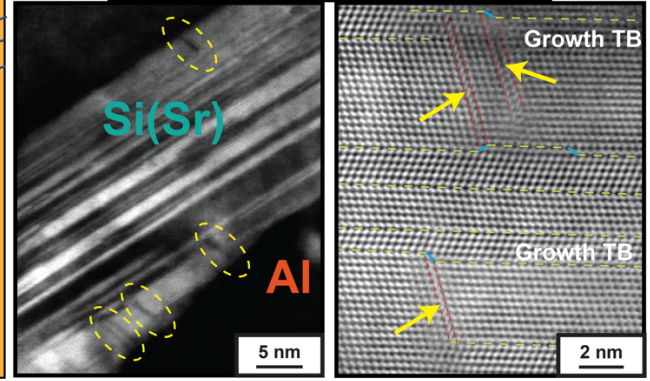

## 1. Introduction

Rapidly solidified ultrafine eutectic composites have gained research spotlight in the past few years because of their extraordinary mechanical properties [1, 2, 3, 4, 5, 6]. Previous works have revealed two types of microstructures inside the melt pool obtained by laser rapid solidification of Al-Si eutectic alloys – first, a hypo-eutectic type heterogeneous microstructure (even in nominally hyper-eutectic compositions) where dendritic primary Al grains are embedded within nano-scale Al-Si eutectic regions and secondly, a fully nano-scale eutectic structure [7]. In both cases, the morphology of the Si phase within the eutectic was fibrous, resulting from flake to fiber transition due to rapid cooling [8, 9]. Although the as-cast microstructure, evaluated using micromechanical test samples from outside the melt pool, revealed low yield strength and low plasticity due to cracking of Si flakes, simultaneous improvement in strength and plasticity was noted in samples from the laser refined microstructures. The deformation mechanisms involved constrained plasticity resulting in geometrically necessary dislocations (GNDs) that increases strain hardening: first, at the interfaces between primary Al and Al-Si eutectic, and second, in the nano-Al channels in the eutectic [10, 11]. Cross-section transmission electron microscopy (TEM) of the nano-indents revealed frequent cracking in the Si fibers such that directly below the indent tip, the morphology transformed from long fibers to particles [12]. Other nanoscale Al-based composites have also shown periodic cracking of the hard (brittle) phase [13].

Atomistic simulation of the interaction of glide dislocations in Al interacting with flat facets of the Si flakes revealed that the edge dislocations climb up along the Al/Si interface and screw dislocations cross-slip onto the interface plane and are reflected back into the Al phase, implying that the energy barrier for slip transmission across interface is significantly higher than slip reflectance from interface [14]. In microstructures with disparate phases, refinement to nanoscale in conjunction with laminate morphology and crystallographic alignment of slip systems across the interface is shown to favor slip transmission [15]. In the case of Al-Si, the high Peierls barrier associated with the covalently bonded Si phase [16] and high tensile residual stresses that develop along the length of the continuous Si fibers [17] make it more challenging to activate slip prior to cracking. To explore the plastic co-deformability, a fully eutectic microstructure with Si refined well below 50 nm diameter, and reduced aspect ratio is needed. However, in the laser melt pool of hyper-eutectic Al-Si alloys, the dendritic primary Al + Al-Si eutectic heterogeneous morphology is dominant with relatively low volume fraction of fully eutectic morphology, thereby limiting a detailed investigation of the mechanical behavior of fully nano-eutectic morphologies.

In the present study, the laser processing approach was modified in two ways. First, an ultra-fine laser spot of 75 μm diameter was used, to increase the input laser energy density and consequently achieve higher cooling rate [18, 19]. Second, chemical effect (0.2 wt.% Sr alloying) was used in conjunction with laser rapid solidification to modify the Al-Si eutectic morphology. While higher cooling rates can favor growth twins upon solidification, particularly in lower stacking fault energy materials [20], minor alloying with rare earth elements, such as Ce, Ba, Sr, Ca, Na, Eu, [21, 22, 23], regardless of cooling rate, can favor growth twins in Si in the Al-Si eutectic [24, 25] and increase the eutectic colony size [26]. In the case of laser solidification, Sr alloying has been shown to increase the volume fraction of the fully eutectic morphology in the melt pool [27]. As shown in the following sections, the modified processing approach resulted in a unique microstructure with regard to the combination of refinement, reduction in aspect ratio and increase in growth twin density in Si, while also significantly increasing the volume fraction of fully eutectic microstructure in the melt pool. This enabled micromechanical testing over a range of sizes to elucidate the enhanced strength-plasticity combination via deformation mechanisms that promote localized dislocation activity Sr-modified Si and suppress cracking.

## 2. Materials and Methods

### 2.1. *Material and processing*

Sr modified (0.2 wt.%) and unmodified Al - 20 wt.% Si hypereutectic alloys were prepared by vacuum arc melting. A hypereutectic base composition of Al-Si was selected since rapid (non-equilibrium) solidification incurs a shift of eutectic point to the hypereutectic side of the equilibrium phase diagram [28] because of solute trapping effect at higher cooling rate [29]. An exact eutectic composition based on equilibrium phase diagram would result in a hypoeutectic microstructure after laser surface remelting (LSR) [7]. Several rectangular pieces with dimensions 20 mm (length: L) × 10 mm (width: W) × 5 mm (thickness: T) were cut out of the as-cast alloys received in button form. The top surfaces were ground with 120-grit SiC sandpaper in order to ensure maximum laser absorption and uniform melt pool microstructure throughout the scanning track [7, 26]. LSR was conducted on the ground surface with laser power, spot diameter, and scanning velocity of 200 W, 100 mm/s, and 75 μm, respectively, using open additive PANDA machine, which is traditionally used for laser powder bed fusion. A stainless steel block was used as heat sink during the laser processing. Melt pool was obtained on the specimen cross-section, as displayed in figure 1a, and metallographically polished for characterization.

### 2.2. *Microstructure characterization*

For optical microscopy (OM), Nikon Metallograph was used. Scanning electron microscopy (SEM) was done with secondary electron imaging mode in TFS Helios 650 NanoLab, using accelerating voltage and beam current of 2 kV and 100 pA, respectively. All the TEM samples were lifted out and thinned by focused ion beam (FIB) in the TFS Helios 650 NanoLab. For standard STEM imaging, STEM based energy dispersive X-ray spectroscopy (EDX), and HR-STEM, TFS Talos F200X G2 was used in 200 kV mode. Aberration corrected 300 kV instrument TFS Spectra (S)TEM was used only for atomic scale imaging. Atom probe tomography (APT) was done in Cameca LEAP 5000 XR Atom Probe, for which the specimens were fabricated by FIB in TFS Helios G4 Plasma FIB UXe.

### 2.3. *Micromechanical testing*

Micromechanical testing, in terms of nanoindentation and micropillar compression, were carried out in Hysitron 950 Triboindenter manufactured by Bruker. Load controlled nanoindentation experiments were performed using high load Berkovich indenter with peak load of 100 mN and loading, holding, and unloading times of 5s, 2s, and 5s, respectively. For calculation of hardness and modulus, a total of 10 experimental load-displacement data were averaged to ensure a decent measurement accuracy. Micropillars of ~6.5 μm diameter and ~14 μm height were fabricated using TFS Helios 650 NanoLab using FIB. Micropillar compression tests were done using a 10 μm flat high load diamond indenter at a strain rate of 0.2%/s and in displacement controlled mode with a maximum displacement corresponding to 30% nominal strain (i.e., 0.3 times the initial pillar height). Stress and strain values were converted from the load-displacement data and measured pillar dimensions. At least three data sets were considered and averaged for measurement accuracy.

## 3. Results

### *3.1. As-processed microstructure*

Single scan LSR was done on the surface of Sr microalloyed as-cast Al-Si hypereutectic composite as schemed in figure 1a, which results in a melt pool on the specimen cross-section with a large volume fraction of fully eutectic structure as observed in the OM image (figure 1b). SEM images of the melt pool (figure 1c) displays that the microstructure is a few hundred times more refined than its as-cast counterpart (figure 1e). As-cast microstructure comprises micron-scale Si fibers dispersed over Al grains. A heterogeneous structure with dendritic primary Al grains embedded within the ultrafine eutectic was observed along the edge of the melt pool (figure 1d). The ultrafine eutectic has micron-scale Al colonies in which nano-scale Si fibers are finely distributed in a specific orientation. Appearance of the fibers depends on their orientation with respect to the optic axis, i.e., electron beam. Different fiber appearances in different colonies have been shown in figure 1c1-1c3, and their orientation with respect to optic axis as well as deformation axis (both are same) is diagrammed in figure 1f. Similar characterization on unmodified Al-Si eutectic is presented in the supplemental document (section 1).

An attempt has been made to qualitatively correlate orientation of Al colonies with Si fiber orientation in the melt pool of Sr modified eutectic by electron back-scattered diffraction (EBSD) imaging. Figure 2a demonstrates inverse pole figure (IPF) of a region with fully eutectic structure, where several colonies and sub-colonies could be spotted. Following standard practice [30], we are denoting two different orientations as colonies (or grains) if the misorientation angle of the boundary separating them exceeds at least 15°. For lower misorientation angle of the boundaries, we are treating the adjacent orientations as sub-colonies (or cells within colonies). Si fiber appearances, i.e., orientations consistently change across different colonies, whereas the change is trivial over the sub-colonies. This observation evinces the possibility of change in fiber orientations primarily because of relative rotation of Al grains assuming habitual growth direction of the fibers. Considerable anisotropy for uniaxial compression behavior of the same region is observed from Taylor map in figure 2b. Moreover, fiber orientation difference is also reflected on the mechanical properties of composites [31]. Therefore, it is important to perform micromechanical tests taking a decent number of colonies into account, which has been reflected on the micropillar compression experiments. That said, it is still essential to study the local deformation behavior of specific colonies. For example, since vertically oriented fibers bear maximum loading, they are most important for crack analysis. Therefore, specific colonies with vertical fiber orientations are of great interest for performing post-loading microstructure analysis. We followed this approach while doing the nanoindentation experiments. In other words, we tried to avoid artifacts like colony or sub-colony boundaries underneath the indent to investigate the post-deformation microstructure more precisely. Figure 2c contains legends of IPF and Taylor map.

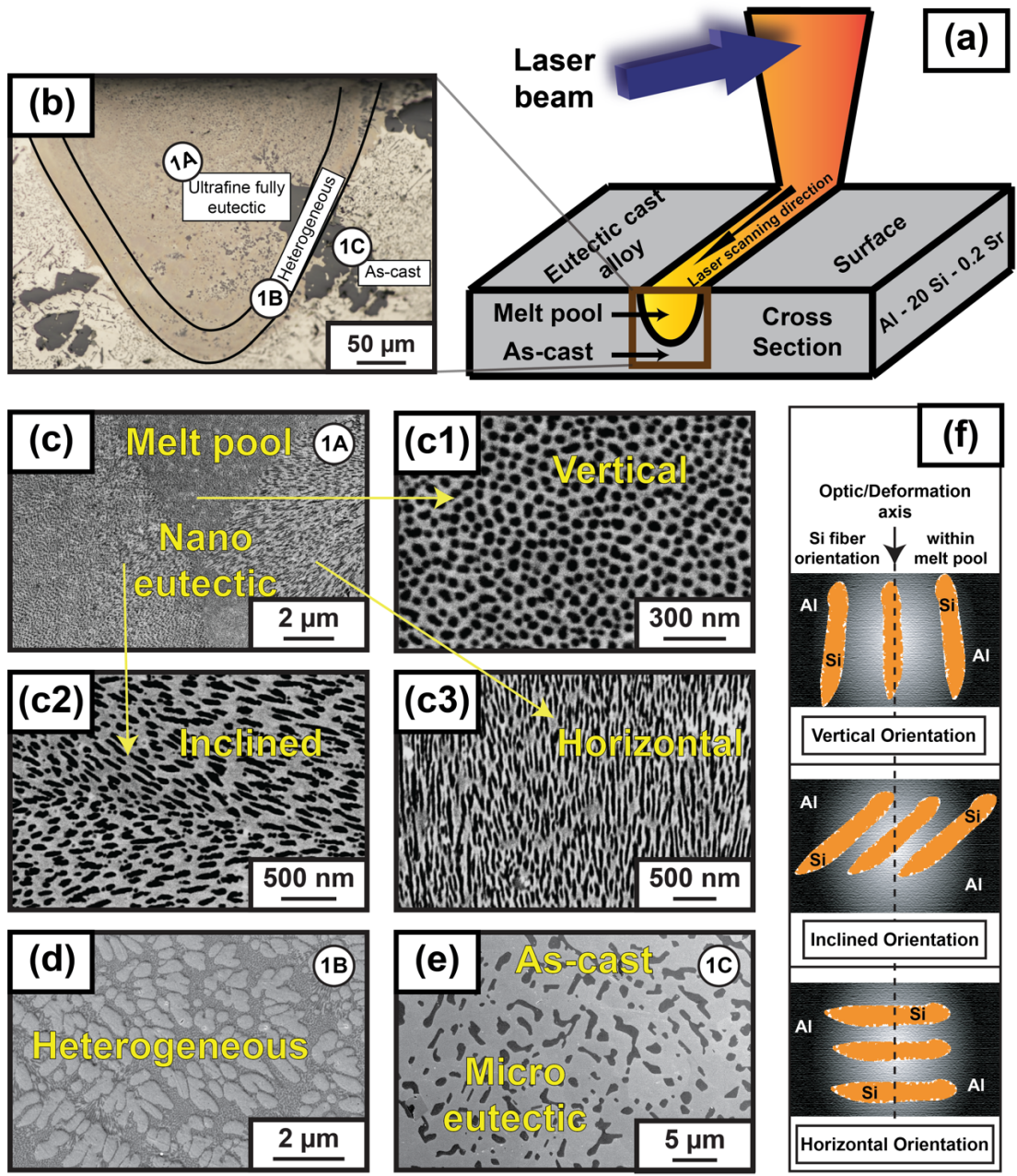

Figure 1: (a) Schematic of the laser surface remelting (LSR) process, (b) optical micrograph (OM) showing the melt pool formation on the specimen cross-section and its surroundings, (c) scanning electron micrograph (SEM) fully nano-scale eutectic region [1A] within the melt pool of Sr modified Al-Si showing different orientation of the hard Si fibers (c1-c3), (d) heterogeneous eutectic along the edge of the melt pool [region 1B], (e) as-cast micron-scale base eutectic composite microstructure outside the melt pool [region 1C], (f) schematic of different fiber orientations with respect to the deformation axis.

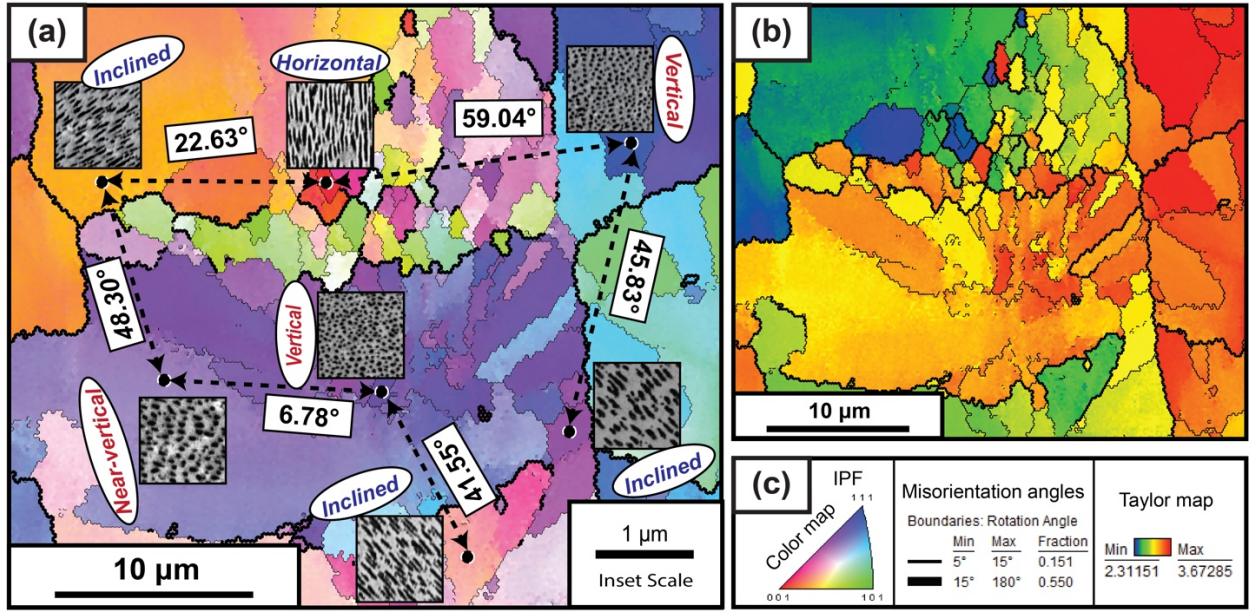

Figure 2: (a) IPF manifesting different orientation of the Al grains (colonies) across a part of the melt pool. Fiber orientation changes from colony to colony, suggesting that the orientation difference across a colony boundary (i.e., misorientation boundary) is possibly because of relative rotation of the colonies (assuming similar growth direction of most of the fibers within a particular colony). (b) Taylor map of the same region displaying mechanical (uniaxial compression) anisotropy of the colonies. Red and blue grains imply hard and soft grains, respectively, under uniaxial compression. (c) IPF color map, misorientation boundary details and Taylor map legend.

The SEM images from figure 1 have been used to measure the fiber dimensions (diameter: d and length: l) and inter-fiber spacing. We are assuming the shape of the fibers to be cylindrical and following k-nearest neighbor method to estimate the inter-fiber spacing. These measurements have been presented and compared with earlier reported values for coarse spot LSR of unmodified alloy [12] as well as with the values obtained from the microstructure of fine spot LSR of unmodified Al-Si through figure 3. Average fiber diameter and inter-fiber spacing values for the Sr modified eutectic are found to be $31 \pm 6$ nm and $44 \pm 9$ nm, respectively, which are somewhat less than the previous report. It is striking that fiber length has dropped by a large margin in this work, which has reduced the fiber aspect ratio by almost four folds compared to previous report. These refinements are direct consequences of 10x higher cooling rate in this work due to finer laser spot. These measured values are very similar for Sr modified and unmodified eutectics.

Sr modification has led to remarkable increment in twin density as evident by comparing the on-zone bright field HR-STEM images of Si fibers from unmodified (figure 4a) and Sr modified (figure 4b) nano-eutectic. Twin thickness and spacing in the modified eutectic Si fiber are in the ranges of 0.5 to 1 nm and 2 to 3 nm, respectively. These nano-twins are mostly unidirectional. The enhanced twin density could be attributed mainly to Sr segregation preferentially in Si fibers (refer to the atom probe tomography analysis in section 4 of the supplemental information) and consequent IIT mechanism. Whereas higher undercooling has a subtle effect on promoting twinning [20], which is reflected by moderate twin density in unmodified fibers. As the effect of Sr modification is mostly manifested in Si fibers, from now on, the fibers in unmodified and Sr modified nano-eutectics will be denoted as Si and Si(Sr), respectively.

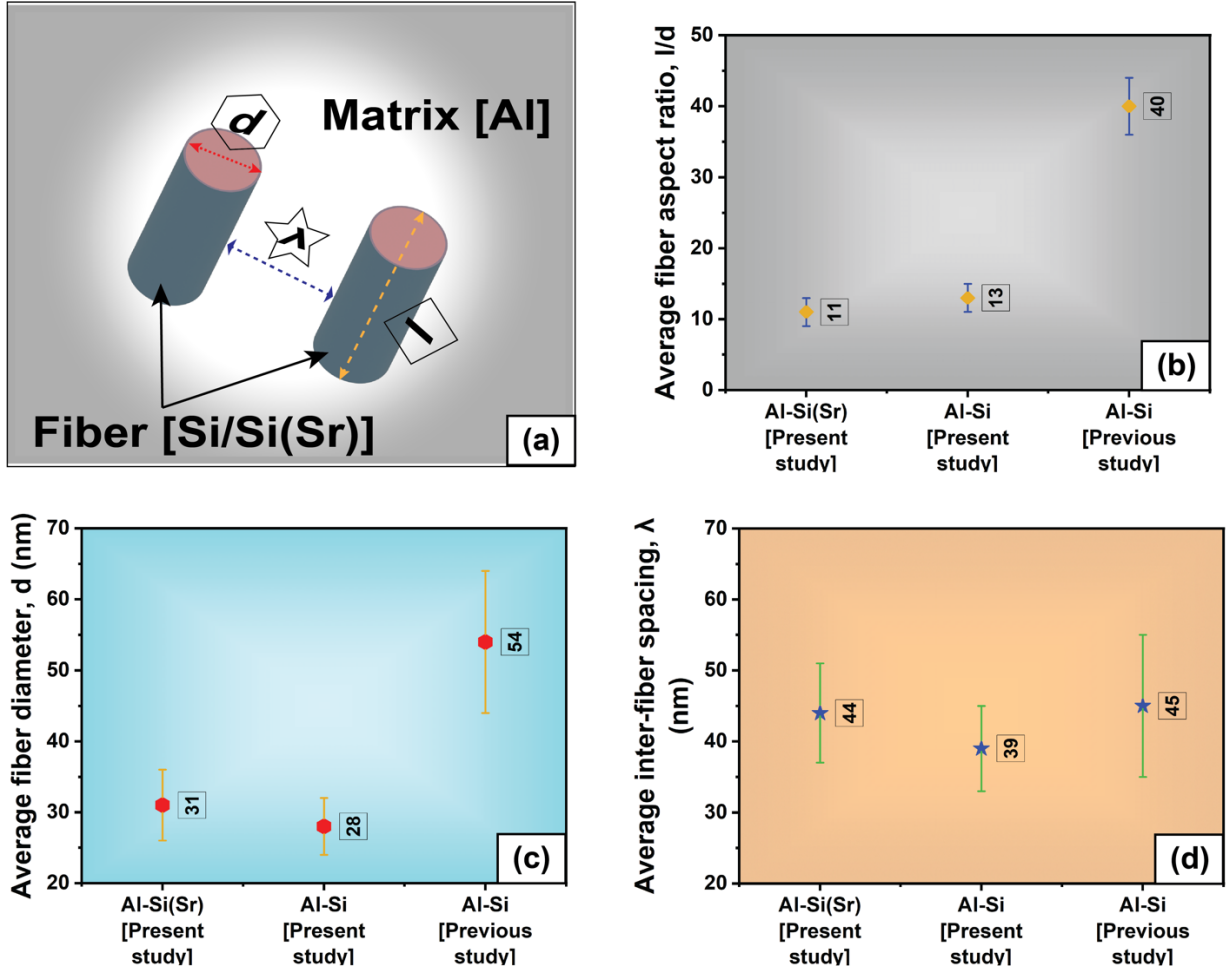

Figure 3: (a) Schematic of the measurement methodology. Comparison of (b) average fiber aspect ratio (length/diameter), (c) average fiber diameter, and (d) average inter-fiber spacing between fine spot LSR of Sr modified Al-Si eutectic (this study), fine spot LSR of unmodified Al-Si eutectic (this study) and coarse spot LSR of unmodified Al-Si eutectic (previous study) [10].

In summary, there are three salient features of the as-processed, i.e., laser rapid solidified microstructure:

        (i) immensely ultrafine Si fibers and inter-fiber spacing,
        (ii) huge density of nano-twins in Si fibers, and
        (iii) segregation of Sr and Al atoms within Si fibers.

Influence of these attributes on the deformation behavior has been addressed in the discussion section of the article.

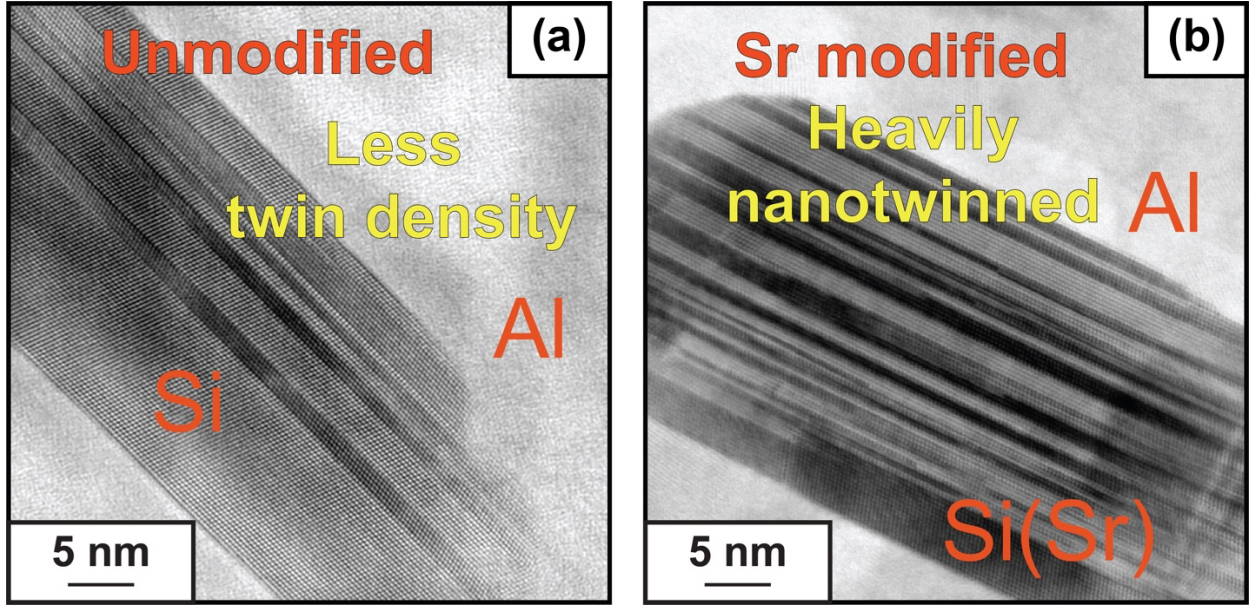

Figure 4: Bright field HR-STEM images depict (a) relatively less twin density in unmodified and (b) high twin density in Sr modified Si nano-fiber. This higher twin density could be attributed to impurity induced twinning as a result of preferential segregation of Sr atoms in the Si fibers.

*3.2. Micromechanical properties*

Nanoindentation experiments were targeted at different colonies having definite fiber orientations with respect to loading axis. Corresponding load-displacement plots and indent impressions used for hardness and elastic modulus calculation can be found in supplemental information (section 6). It should be noted that the calculated hardness values following standard area of contact method [32] match quite well with the machine generated tip-calibrated values. Penetration depth, hardness, and Young's modulus values have been reported in table 1. Depending on the depth of indent (D ~1 μm), we can assume that the plastic zone under the indent is extended to minimum three times of D (note: we are considering the minimum depth of plastic zones in order to safely stay within plastic zones while characterizing microstructures under the indents) [33], i.e., minimum ~3 μm. We have characterized the deformed microstructures under the indents well within this distance limit.

Table 1: Nanoindentation obtained mechanical properties of Sr modified and unmodified nano-eutectics

| Nano-eutectic | Indentation depth | Hardness | Young's modulus |
|---|---|---|---|
| Al-Si(Sr) | 1100 ± 100 nm | 2.7 ± 0.4 GPa | 96 ± 4 GPa |
| Al-Si | 1000 ± 30 nm | 2.9 ± 0.1 GPa | 99 ± 3 GPa |

Higher standard deviation in the values for Sr modified eutectic is an indication of strong anisotropic response. As Sr addition increases colony size, plastic zones under the indents are almost accommodated by a single colony. That is why, mechanical properties have varied considerably from colony to colony, since the nanoindentation tests were confined in particular colonies. Therefore, hardness of specific Al colonies along with fiber orientations with respect to indent axis have intrigued the anisotropy. This anisotropy is not observed in the unmodified counterpart because of smaller colony sizes.

Micropillars were fabricated over decent number of colonies to avoid local anisotropy. Figure 5a presents stress-strain plot obtained from micropillar compression and table 2 quantifies the data. We have considered the stress-strain data until that point where the pillar retained its shape and deformed uniaxially. Instability point can be assessed from the stress-strain plots too, as there is a large fluctuation in the direction of the plot. As can be seen from the pillars compressed up to a 25% of nominal strain, there are traces of slip on the Al-Si(Sr) pillar body without any prominent sign of brittle failure (figure 5b). On the other hand, the unmodified Al-Si pillar is cracked and has not maintained its cylindrical shape (figure 5b), which is also reflected by the stability of the Al-Si pillar until a lower plastic strain compared to Al-Si(Sr) pillar. It is noteworthy that we are not quantifying the plasticity of these microstructures distinctly, as the prediction of ductility through compression experiments is often misleading [34]. Rather, the aim is to compare the mechanical stability of different microstructures in terms of applied plastic strain.

Table 2: Mechanical properties of Sr modified and unmodified nano-eutectics obtained from micropillar compression experiments.

| Nano-eutectic | True compressive stress | True plastic strain until mechanical stability |
|---|---|---|
| Al-Si(Sr) | 840 MPa | 26% |
| Al-Si | 820 MPa | 21% |

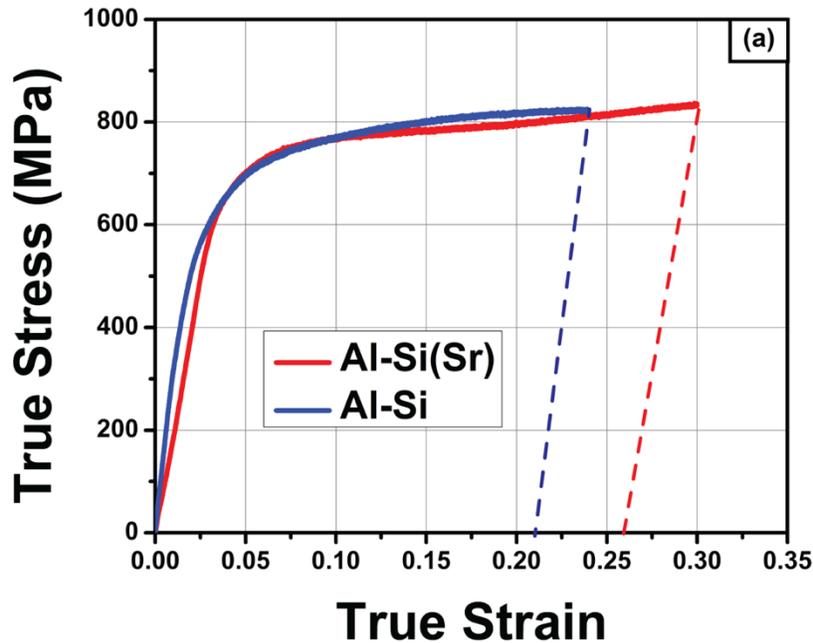

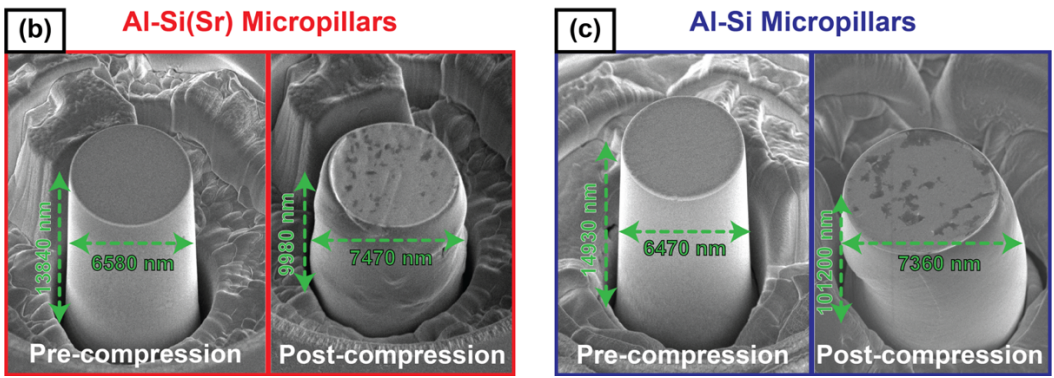

Figure 5: (a) Stress-strain plots attained by micropillar compression of Al-Si(Sr) and Al-Si. Pre and post-compression SEM images of the (b) Al-Si(Sr) and (c) Al-Si pillars help understand the stability of each microstructure at same loading conditions.

It is pertinent to mention that hardness, strength, and plastic strain values acquired through the micromechanical tests are way higher for the nano-scale eutectics than their as-cast counterpart. To get an estimate of the mechanical properties and deformed microstructure of as-cast microstructure, one is referred to the supplemental information (section 7) of this article. We have also noted considerable improvements in hardness, compressive strength, and mechanical stability in this work compared to previously reported values for coarse-spot LSR processed unmodified Al-Si [10, 12]. It is, therefore, necessary to characterize deformed microstructure to find out the exact contributing factors.

### 3.3. *Deformed microstructure*

Since vertically oriented fibers with respect to deformation axis bear maximum load compared to other orientations, it is most likely that any crack formation would be manifested in those fibers first during loading [12]. Cross-section TEM samples from the indented regions of specific colonies having near-vertical orientation of fibers with respect to loading axis in Al-Si(Sr) nano-eutectic were prepared to look at the fibers along their length and also to observe overall defect structure in the plastic zone. STEM-EDX does not indicate any sign of fiber fracture even under the closest proximity of the indent in figure 6a. Bright field and dark field STEM images provide a closer view to the long parallel crack-free fibers in figure 6(b1-b2). It was striking indeed to spot planar faults within Si(Sr) fibers near the Al/Si(Sr) interfaces, which are marked with yellow arrows. These faults were not present prior to deformation. EDX partitioned transmission Kikuchi diffraction (TKD) maps underneath the indent helped assess the overall misorientation profile of each phase. It is quite evident from the IPF given in figure 6c1 that the entire part of the plastic zone we are dealing with almost belongs to a single colony and EDX map of Al and Si(Sr) of the same region is shown in figure 6c2. Figure 6c3 and figure 6c4 present Kernel average misorientation (KAM) maps of Al and Si(Sr), respectively. Although intracolonial misorientation was negligible before deformation, it has remarkably increased after indentation. It is possible to directly correlate the KAM with dislocation density ($\rho$), where the 0° (black) to 5° (red) KAM scale coincides with 0 to $1.5 \times 10^{14}$ $m^{-2}$ of $\rho$ scale [35]. Just under the indent, dislocation density in Al is expectedly very low as a result of rapid dynamic recovery. Interestingly, there is pronounced misorientation in the KAM map of Si throughout, which is quantitatively of higher order than the intra-colonial misorientation prior to deformation. Therefore, higher misorientation in Si might have been originated as a result of plastic deformation induced defects. In order to confirm this hypothesis, we have carried out atomic scale HR-STEM of Si(Sr) fibers in the deformed microstructure.

Atomic resolution STEM images of the Si(Sr) fibers underneath the indent exhibit formation of SFs quite often due to local shearing near the interfaces (figure 7a) which extend until a growth twin boundary (TB) (figure 7b). It is to be noted that the post-deformed microstructure of the Si(Sr) fibers is fairly different from that of the as-remelted fibers (refer to section 11 of supplemental document for a clear visualization). Partial dislocations nucleate near the interfaces and glide leaving an intrinsic SF behind, which can be visualized in figure 7c. Burgers circuit analysis surrounding the dislocation confirms its Burgers vector to be $(1/6)$ $[1\ \text{-}2\ 1]_{Si}$. Sometimes the fault structure is also prominent even in the center of the fibers as shown in figure 7d. An example of multiple fault driven step formation in $\{1\ 1\ 1\}_{Si}$ planes near the Al/Si(Sr) interface is captured in figure 7e1, which is magnified in figure 7e2. This fault driven phenomenon was restricted at the very first growth TB close to the interface.

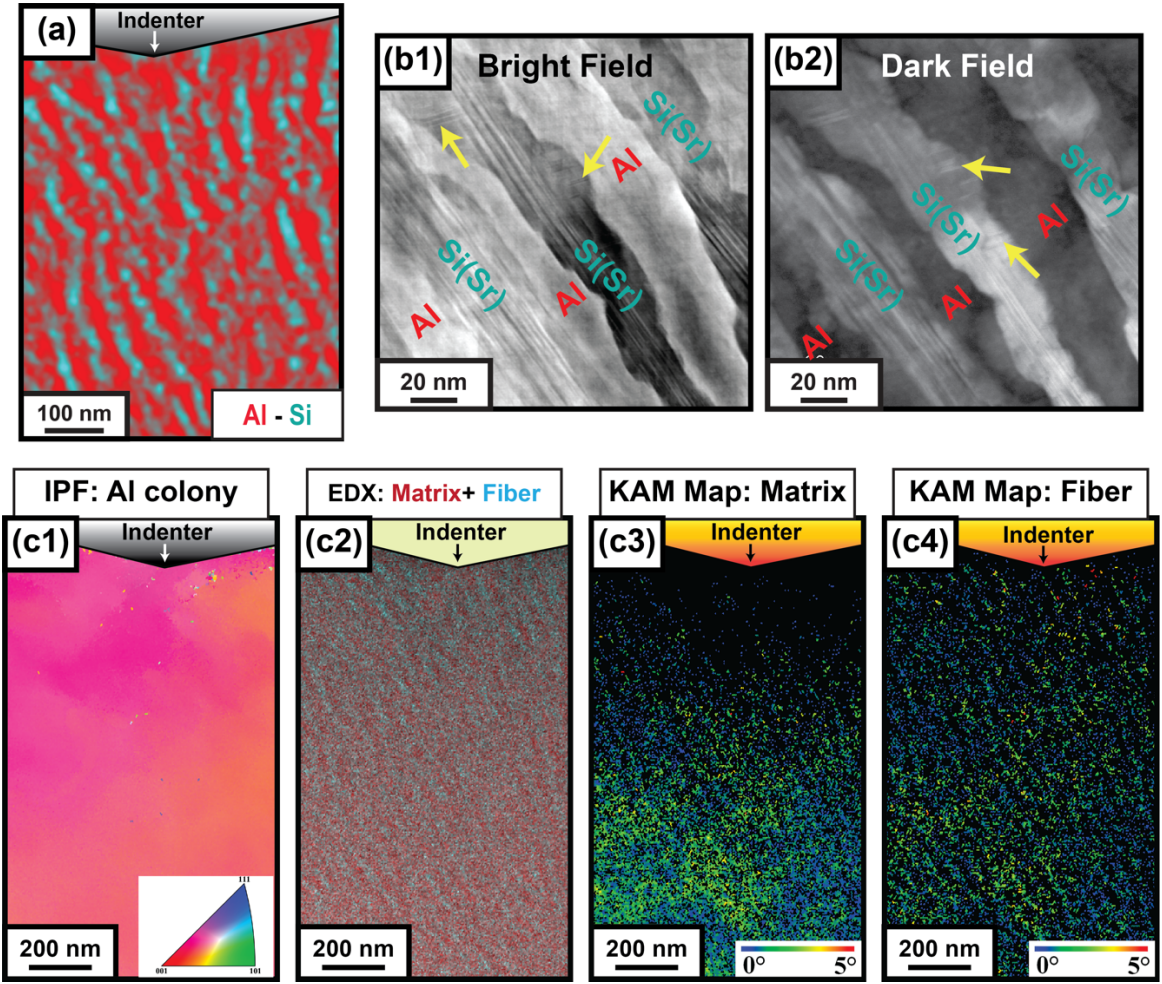

Figure 6: STEM-EDX map of Al and Si underneath the indent envisaging no crack formation in the Si(Sr) fibers, (b1-b2) bright and dark field STEM images of crack-free long parallel fibers with traces of planar faults near the interfaces. TKD micrographs of the same indented area is demonstrated in terms of (c1) IPF that confirms single colony across the region of interest within plastic zone. (c2) Simultaneous EDX map shows near vertically oriented nano-fibers (blue) in Al matrix (red). KAM maps point out misorientation profiles in (c3) matrix (Al) and (c4) fiber (Si) phases within that single colony.

While crack-free Si(Sr) fibers exhibit deformation instigated planar faults, figure 8 shows high dislocation density in narrow Al channels. Figure 8(a1-a3) portrays single dislocation arrays in narrow Al channels in between elongated Si(Sr) fibers. Figure 8(b1-b3) presents the dislocation structure in plastic zone of another specimen, where fibers are horizontal with respect to indent axis and are perpendicular to the plane of the foil sample. It gives a view of single dislocation networks in Al channels distributed throughout the plastic zone connecting the heads of the Si(Sr) fibers. The primary cause of huge strain hardening in nano-scale eutectic is formation of single dislocation arrays in the ultrafine Al channels, while the Si fiber fracture is delayed to a competent strain level because of ultrafine fiber dimensions [10, 11]. In the as-cast eutectics, coarse Al phases in between Si fibers undergo normal strain hardening similar to bulk Al, where the strengthening is limited by easy cross-slip [36]. Stress concentrations develop in hard Si fibers, which crack beyond a certain stress, causing failure of the entire microstructure.

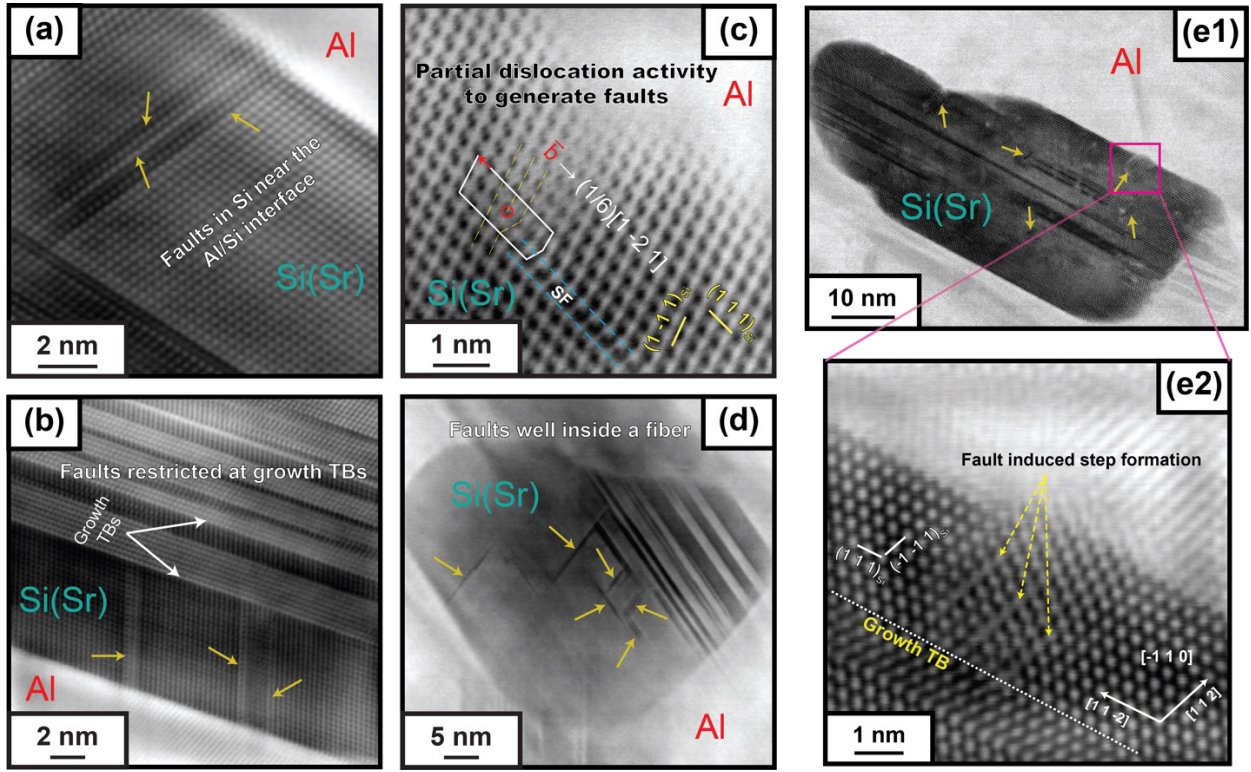

Figure 7: (a) Deformation (nanoindentation) induced SFs in Si(Sr) fibers near the Al/Si(Sr) interface. (b) These SFs get restricted at the growth TBs. (c) SFs are originated as a result of partial dislocation mobility in $\{1\,1\,1\}_{Si}$ planes having Burgers vector along $<1\,1\,2>_{Si}$ directions. (d) Sometimes the planar faults are also present well within the fibers and get stuck at the growth TBs or SFs. (e1) An example of multiple planar faults in Si(Sr), (e2) which at higher magnification reveals multiple SF induced step formation near the interface in $\{1\,1\,1\}_{Si}$ planes. This phenomenon breaks off at the first growth TB the faults interact with.

The compressed Al-Si(Sr) micropillars were characterized using transmission electron microscope (TEM) to investigate if the deformation induced planar fault structure in Si still prevails. A heavily deformed part of the compressed micropillar lift-out is shown in figure 9a. The inset figure is a low angle annular dark field (LAADF) STEM image from the white outlined region that unveils crack-free long Si(Sr) fibers with a specific orientation. Even though there is no relative rotation or crack noticed in the fibers in that particular region, off-zone axis selected area diffraction pattern (SADP) of the same region in figure 9b indicates splitting of Si peaks, which could be because of high intra-fiber misorientations incurred by deformation [37]. Further HR-STEM analysis clarifies high planar fault density within Si fibers (figure 9c) after pillar compression. The key characteristic of these deformation faults is they are shorter (a few nanometers) than the growth induced faults or TBs. A specific fault, outlined by white square, is chosen for high magnification imaging. Corresponding fast Fourier transform (FFT) pattern is added as an inset image, which helped determine crystallographic directions in figure 10(a1-a2).

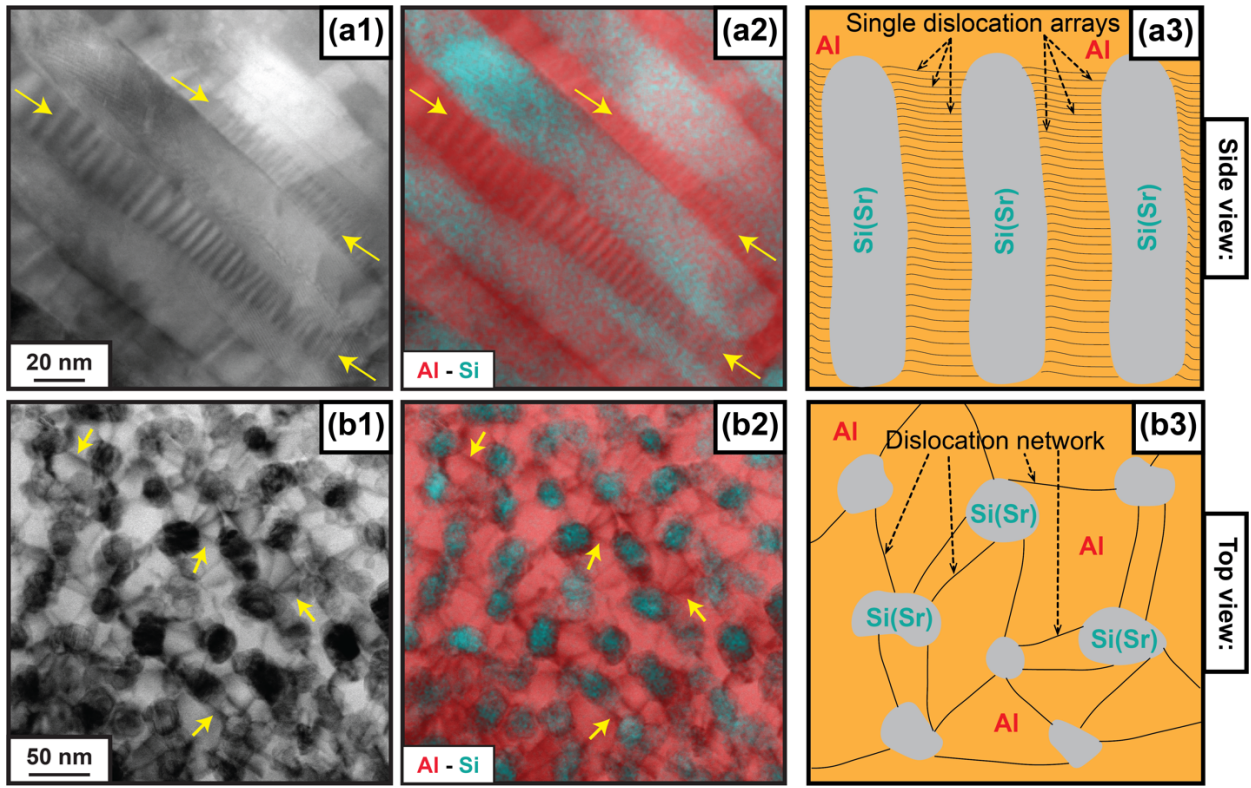

Figure 8: On-zone bright field STEM images pointing out (a1) single dislocation arrays in narrow Al channels between Si(Sr) fibers and (b1) a different orientation of fibers help understand these dislocations are prevalent throughout the plastic zone. Respective EDX maps of (a1) and (b1) have been superimposed in (a2) and (b2). (a3) and (b3) schemes the dislocation profiles as seen in (a1) and (b1), respectively. In 3-D, one can imagine (a3) the first schematic as the side view, which displays dislocation arrays, while (b3) the second schematic can be imagined as the top view, where dislocation networks in Al are visible throughout connecting the heads of the fibers.

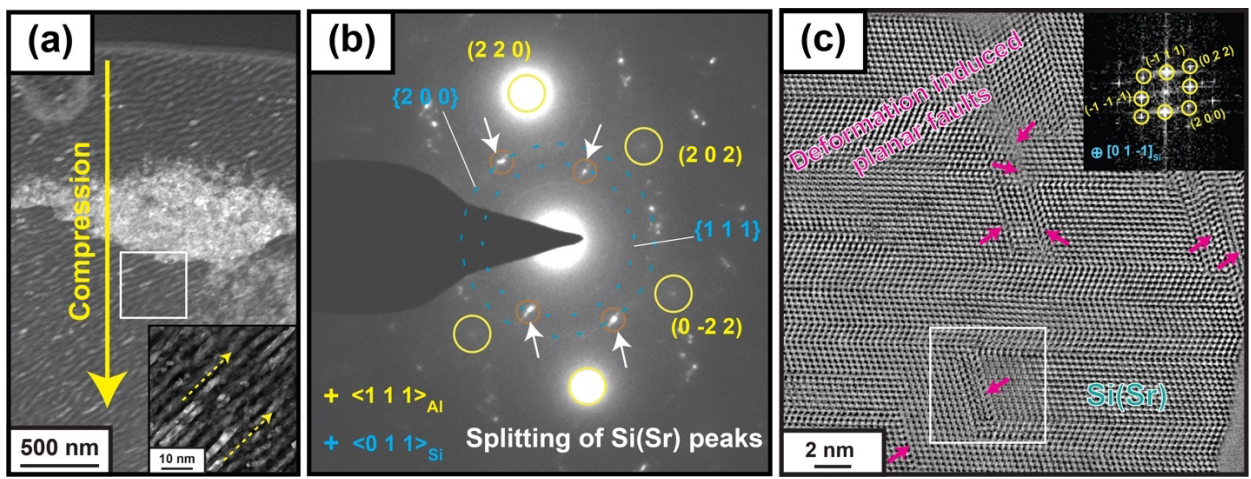

Figure 9: (a) A part of the compressed micropillar used for STEM imaging. Inset image is a LAADF STEM micrograph of the white outlined region that exhibits crack-free elongated Si(Sr) fibers with a specific orientation. (b) SADP of the same region reveals splitting of Si peaks giving a hint of plastic deformation

of Si(Sr). (c) Bright field HR-STEM of a particular Si(Sr) fiber with high density of deformation induced short range SFs (marked with purple arrows). FFT pattern of the white outlined region is placed as an inset image. This outlined area having a deformation SF is targeted for atomic scale HAADF STEM imaging.

Figure 10a1 presents an atomic scale high angle annular dark field (HAADF) STEM image of a Si(Sr) fiber from the deformed pillar that shows formation of a SF on $(1\ 1\ 1)_{Si}$ plane along $[2\ -1\ -1]_{Si}$ direction originated from a step on a growth TB (white dotted encircled). It is important to notice that in the as-processed Si(Sr) fibers (figure 10b), i.e., prior to deformation, there exist some prominent growth induced features, viz., high unidirectional twin density with TB directions along $<1\ 1\ 2>_{Si}$, step formation on the TBs (twinning/detwinning), and presence of partial dislocations at those steps as well as at the extreme points of TBs which are marked with blue circles. However, any short range SFs confined within the growth TBs have not been detected. Therefore, the characteristic SF highlighted in figure 10a1 could be attributed to shear deformation. Needless to mention, we are comparing the images taken from same zone axis. Figure 10a2 is the magnified image of figure 10a1, which focuses on the partial dislocation activities across the deformation SF. The blue encircled partials are generated during growth as mentioned earlier and are responsible for phenomena like growth twin formation, step formation on growth TBs. The partials marked red are governed by deformation and these extended partials gave rise to the SF (pointed up by white dotted lines). It is also evident that the glide of the partial below the SF has been restricted by another growth TB. Deformation generated partial dislocation activities in Si at room temperature is not very commonly reported. Therefore, it is important to determine the Burgers vector of the partials. The partial dislocation formation site at the growth TB step has been reconstructed by a molecular statics model in figure 10a3. It confirms the deformation induced partial (A) is a mixed dislocation in nature with a Burgers vector ($b_A$) of $(1/6)[2\ -1\ -1]_{Si}$.

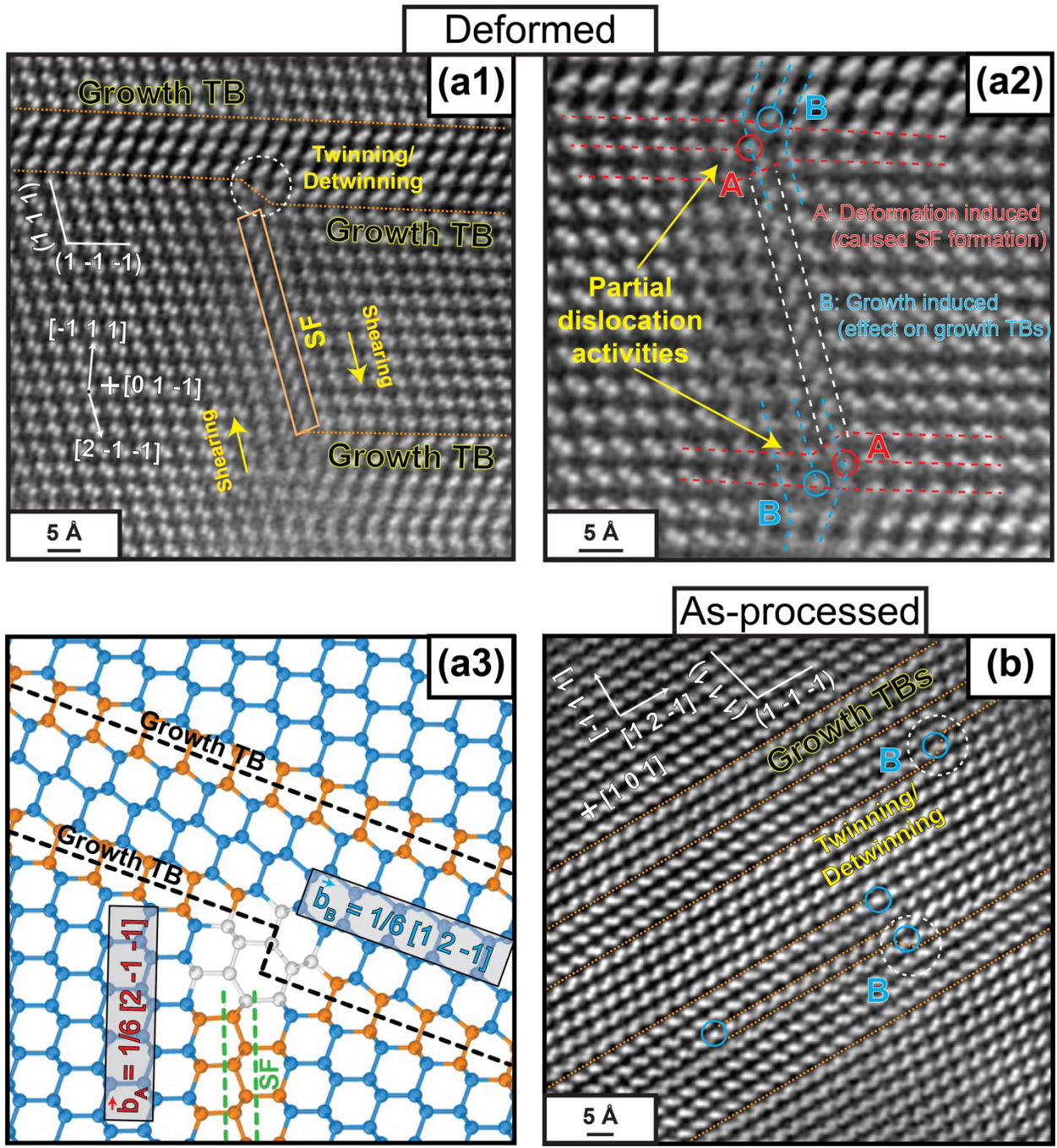

Figure 10: Atomic resolution HAADF STEM images of Si(Sr) fibers highlighting (a1) formation of a deformation induced SF at the step formed on a growth TB and its propagation until a barrier created by a second growth TB and (a2) the partial dislocation activities across the SF. (b) As-processed HAADF STEM micrograph from the same zone axis is used to distinguish between the growth driven and the deformation incurred phenomena. Comparing the images from post and pre-deformed Si fibers, it can be concluded that deformation results in partial dislocation nucleation at the step on the growth TB, which glides leaving a SF behind, until its movement is hindered by another growth TB. (a3) Molecular statics reconstruction of the partial dislocation nucleation site to determine its Burgers vector.

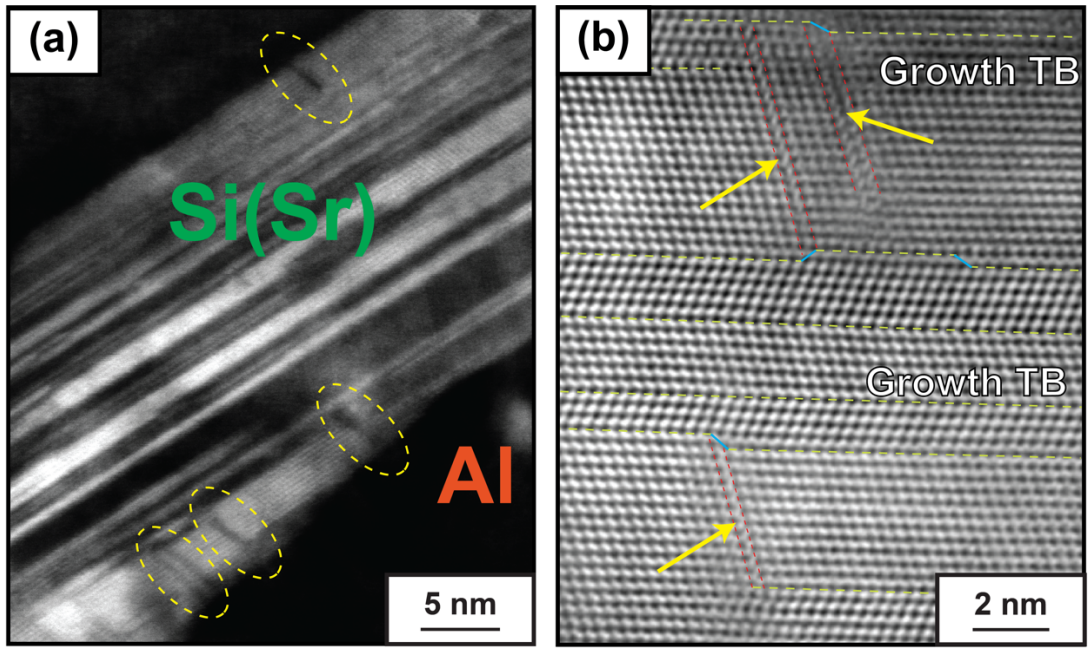

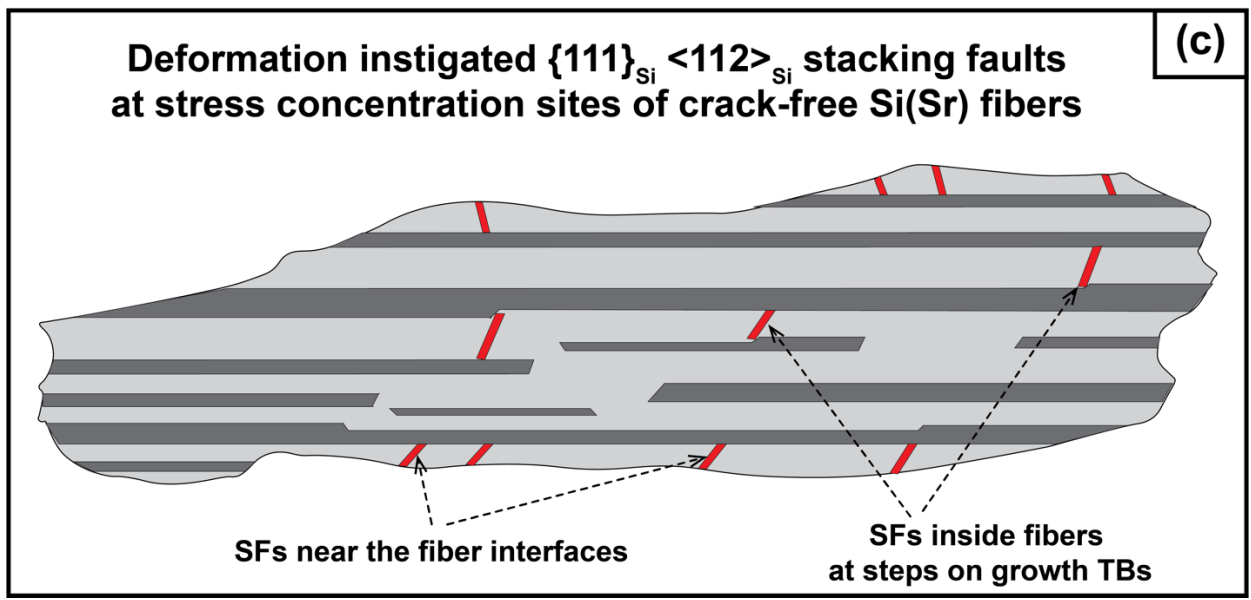

Figure 11: Si(Sr) fibers from compressed micropillar having deformation instigated faults (a) near the interfaces and (b) at the steps on growth TBs. (c) A schematic illustrates these fault formations in an elongated crack-free Si(Sr) fiber.

The deformation induced partial dislocation activities and consequent SF formations at the fiber interfaces as well as at the steps on growth TBs have been summarized in figure 11(a-b) and illustrated in figure 11c.

Similar characterizations were done on the deformed regions from unmodified Al-Si alloy. Figure 12 a portrays a STEM-EDX map of Al and Si under the indent. Si fibers immediately under the indent (zone 1) are fractured and no deformation induced planar faults have been marked, as presented in figure 12b. The region below zone 1, denoted as zone 2, has Si fibers that are cracked, but some typical deformation faults have been spotted (figure 12c). Zone 3, below zone 2, is far from the indent but within the plastic zone, exhibits crack-free long fibers with sign of post-deformation planar faults.

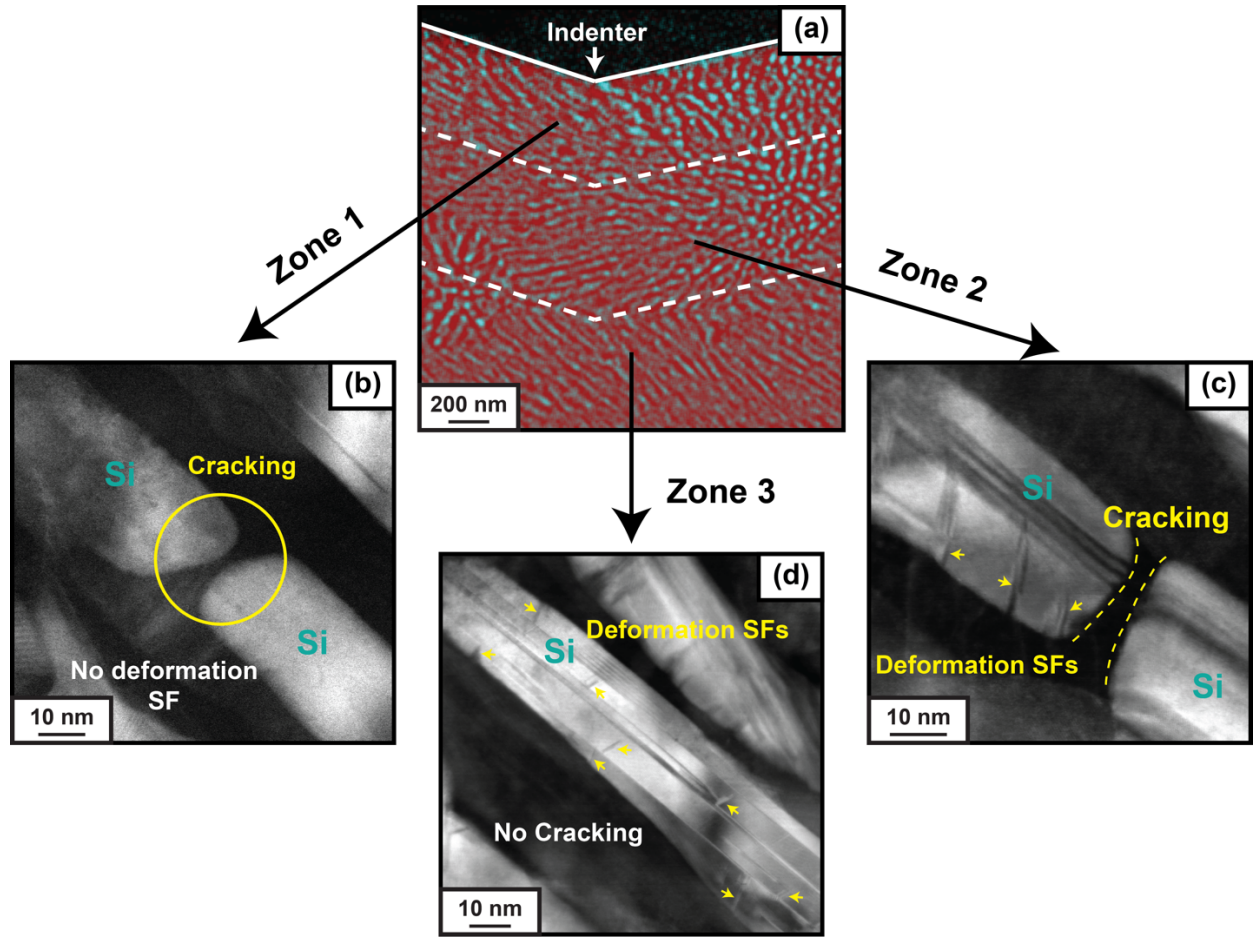

Figure 12: (a) STEM-EDX map under the indent of unmodified Al-Si nano-eutectic, displaying three characteristic zones. (b) In zone 1, Si fibers crack easily with no sign of deformation induced faults. (c) In zone 2, post-deformation faults can be spotted in cracked fibers. (d) In zone 3, fibers are crack-free with a decent density of deformation instigated faults.

The same phenomenon has been observed in case of the compressed micropillar. Figure 13a shows a compressed micropillar, where area 1 is heavily deformed and area 2 has effectively lower deformation effect. In area 1, Si fibers are often cracked as indicated by Si partitioned STEM-EDX in figure 13b and confirmed by LAADF and bright field STEM images in figures 13d1 and d2, respectively. Furthermore, characteristic deformation generated faults were not noticed as revealed by figure 13d3. However, in area 2, where the extent of deformation is comparatively less, post-compression SFs in the fibers become prominent (figure 13e1). These faults are mostly nucleated at the interfaces (figure 13e2). Some SFs are present within the fibers but restricted at growth TB. However, the density of deformation induced SFs within the Si fibers is markedly less than that in Si(Sr) fibers after micropillar compression.

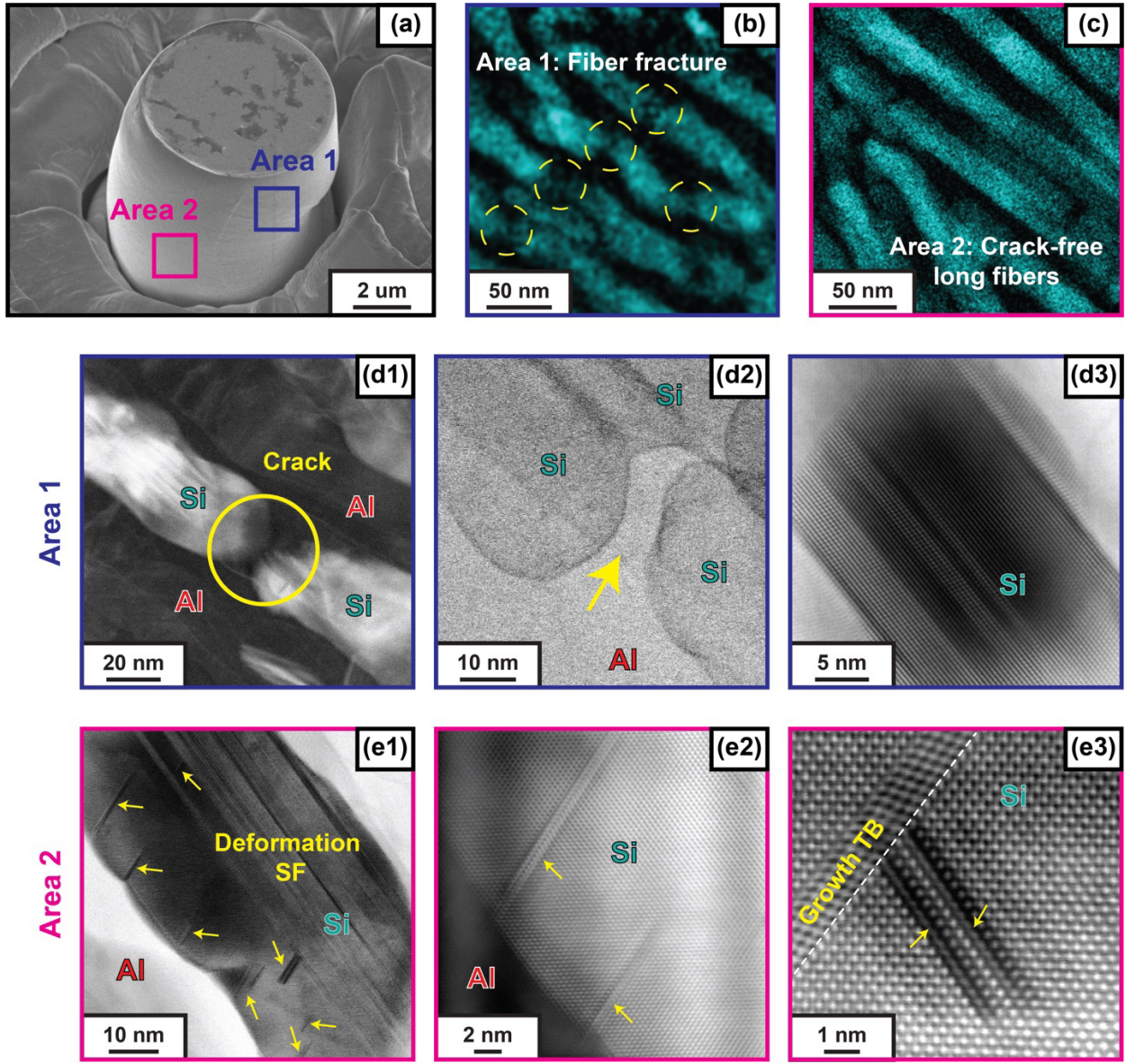

Figure 13: (a) Compressed Al-Si nano-eutectic micropillar, where area 1 is highly deformed and area 2 has lower extent of deformation. Si partitioned STEM-EDX maps hint (b) fiber fracture in area 1 and (c) crack-free fibers in area 2. (d1) LAADF and (d2) bright field STEM micrographs point out different fiber cracking phenomena in area 1, (d3) where deformation induced cracks have not formed in the fibers. (e1) Bright field STEM micrograph highlighting deformation induced SFs in Si fibers in area 2. HR-STEM images reveal the SFs (e2) near fiber interface and (e3) inside the fiber.

## 4. Discussion

This section is intended to discuss the sources of ultrahigh strength, different attributes that have suppressed fiber fracture in Al-Si(Sr) nano-eutectic and enabled plastic co-deformation of the disparate phases, and possible deformation mechanism behind huge strengthening along with compressive stability. In each section, primary emphasis has been given on the influence of as-remelted microstructure parameters on mechanical behavior.

## 4.1. Source of ultrahigh strengthening

It is quite clear that the primary strengthening comes from formation of GNDs in Al matrix near the Al/Si interfaces for both of the modified and unmodified nano-eutectics. Thickness of the narrow Al channels in between the Si fibers, being comparable to the lengths of single dislocations, allow single dislocation arrays to form, as described in figure 8. Strengthening from these dislocation arrays are very similar to confined layer slip (CLS) model [38] and can be termed here as confined channel slip (CCS). In order to estimate the strengthening from GND formation ($\alpha'\mu b\sqrt{[\frac{2\sqrt{3}\varepsilon_p}{bh}]}$) followed by CCS mechanism ($M\frac{\mu b}{8\pi h'}\frac{(4-\nu)}{(1-\nu)}\left[ln\frac{\alpha h'}{b}\right] - \frac{f}{h}$), equation 1 can be used:

$$\sigma_{flow} = \alpha'\mu b\sqrt{[\frac{2\sqrt{3}\varepsilon_p}{bh}]} + M\frac{\mu b}{8\pi h'}\frac{(4-\nu)}{(1-\nu)}\left[ln\frac{\alpha h'}{b}\right] - \frac{f}{h} \qquad (1)$$

Here M is the Taylor factor of Al, the average value of which is found to be ~3 from the Taylor map (figure 2b). The values of other parameters used here are in line with a relevant previous work [10] and are as follows: $\mu$ (shear modulus of Al) = 26.1 GPa, b (Burgers vector) = 0.286 nm (calculated for full dislocations on $\{1\ 1\ 1\}_{Al}$ planes as revealed by figure 10b), $\nu$ (Poisson's ratio) = 0.34, $\alpha$ (core cut-off parameter) = 0.6. Layer thickness (h) is same as inter-fiber spacing = 44nm. As revealed from local orientation data across Al/Si interface in a microstructure having vertically oriented fibers, primary glide plane of Al, i.e., $\{1\ 1\ 1\}_{Al}$ is inclined at angle of 30° with measuring direction, which is along vertical axis (see section 10 of the supplemental information for detailed analysis). Therefore, projected layer thickness (h') = h/cos 30° = 50.8 nm. The ratio of (f/h), termed as normalized interface stress, for a layer thickness of ~44 nm is approximately 50 MPa [38]. At a plastic strain of ~25%, the calculated flow strength is 680 MPa, which is distinctly less than the experimental true stress reported in table 2. This difference could be primarily attributed to strengthening from Al colony boundaries. Interestingly, the calculated strength matches fairly with the true stresses obtained from pillar compressions performed in single colonies (this data is provided in section 9 of the supplemental material).

## 4.2. Suppression of Si(Sr) fiber fracture

While the Al channels beside Si fibers shear to produce a large degree of strain hardening, suppression of Si(Sr) fiber fracture delays bulk failure till a large plastic strain. Unmodified Si fibers, however, crack at high stresses, which is reflected in relatively early failure of Al-Si micropillars. Combination of two contributing factors come into action to prevent the fiber fracture: (i) short fiber lengths and (ii) nucleation and glide of partial dislocations during deformation, that relieve stress concentrations.

For a fiber reinforced composite, the maximum stress ($\sigma_{max}$) at the midpoint of an elongated fiber is proportional to the aspect ratio (l/d) and the shear stress in the matrix ($\tau_m$), which could be expressed by equation 2 [17, 39]:

$$\sigma_{max} = (\frac{l}{d}) \times \tau_m \qquad (2)$$

Fiber fracture occurs when $\sigma_{max}$ is equal to the fracture strength of the fiber. Therefore, smaller aspect ratio will make it more difficult for the Si fibers to fracture. Considering the calculated aspect ratio (l/d) in figure 2b and assuming $\tau_m$ as the flow stress obtained by equation 1, $\sigma_{max}$ values are approximately 7.5 GPa and 8.5 GPa, respectively, for Al-Si(Sr) and Al-Si. These values decrease with decreasing deformation level.

Besides short fibers, partial dislocation nucleation and glide also relieve the local stress concentrations, thereby prevent fiber fracture. During deformation, stress concentrations develop in hard Si fibers initially at the Al/Si interfaces [40]. Also, it is experimentally proven that the steps on pre-existing TBs act as stress concentration and partial dislocation nucleation sites upon applied stress [41, 42]. Partial dislocation activities and subsequent SF formation have been marked predominantly at these stress concentration sites.

However, glide of partial dislocations in Si requires glide set activation, which is not commonly reported at room temperature. According to the widely accepted hypothesis, dislocation movement at RT is energetically more favorable along shuffle sets as only one bond is needed to be broken to initiate dislocation glide, whereas for glide set activation, three atomic bonds are required to be broken, which is reflected by higher Peierls energy barrier for glide dislocations at RT [43]. At elevated temperature, typically above brittle-to-ductile transition temperature, glide set activation is possible because of thermal compensation of Peierls barrier [43]. However, dislocations move by formation and migration of kink pairs in covalent materials like Si [44]. Line energy calculations by Ren et al. clearly showed that, only for stresses below $0.01\mu$ (where $\mu$ represents the shear modulus of Si), the free energy required for the formation of kink pairs was lower for a glide partial dislocation than for a shuffle perfect dislocation [45]. On top of that, density functional theory (DFT) calculations have shown that glide-set dislocations have lower core energy after dissociation below that stress level [46, 47]. Once nucleated, we should consider the energetics of their propagation. MD simulations show that Peierls stress for full dislocation motion in Si (~3.7−4.4 GPa) [48] is higher than that for partial dislocations (~3.3 GPa) [49]. Based on the dislocation-transition model in fcc metals, which share the same slip system as Si, it is evident that the critical resolved shear stress (CRSS) for full dislocations is greater than that for partial dislocations in Si at very low stresses [50]. Formation and glide of partial dislocations in unmodified Si fibers in the less deformation-affected regions could be understood from the above literatures. However, at highly deformed regions, the applied stress on Si or Si(Sr) fibers, estimated from equation 2 is way higher than the stress limit predicted by Ren et al. Quite expectedly, the Si fibers exhibit cracking without any prominent sign of deformation driven partial dislocation activities at the highly deformed regions as understood from figure 12 and figure 13. However, the Si(Sr) fibers do not show any evidence of fracture at any point and provide enough evidence of partial dislocation activities. In fact, these partial dislocations in turn prevent fiber fracture by relieving stresses.

Now the question comes, why the partial dislocations could nucleate and glide at very high stresses (i.e., right under the indent of heavily deformed part of micropillar) in Si(Sr) fibers but not in Si Fibers. Firstly, it might be because of very high twin density in Si(Sr) fibers. Steps on the TBs are potential nucleation sites for the partials for fcc materials that have same slip system as Si [41]. Further it can be argued that Sr and Al segregation in Si fibers might have a pivotal contribution in promoting nucleation of partial dislocations under applied stress since existing literatures support that doping causes reduction of kink-pair formation energy due to an electrostatic interaction [51, 52, 53]. In fact, higher Sr segregation also results in greater Al accumulation inside Si fibers [54]. As mentioned before, partial dislocation nucleation in Si is dominated by kink-pair formation, presence of impurity atoms might enhance the possibility of partial dislocation assisted phenomena that we have observed. Moreover, presence of foreign atoms is expected to weaken the covalent bonds in Si [53]. Propagation of the nucleated kinks involves breaking and reforming the bonds. Therefore, weaker the bonds, easier it is for a kink-pair to migrate [53], which reduces the barrier of partial dislocation movement. Because of reduction in kink-pair migration energy, true extent of partial dislocation activities might have been much larger than the assessment based on SF observation, as the trailing partials might have glided too annihilating many of the SFs.

*4.3. Overall deformation mechanism in Al-Si(Sr) nano-eutectic*

Based on the understanding with primary source of strengthening and suppressed fiber fracture, the deformation mechanism could be summarized with two attributes, as illustrated in figure 14. Onset of plastic deformation is through nucleation of GNDs near the soft (Al)/hard (Si) interfaces and formation of single dislocation arrays at the thin Al channels between Si(Sr) fibers, governed by CCS mechanism. Shear deformation of Al has been marked as the first attribute. During deformation, Si fiber fracture is initially suppressed because of their ultrafine length-scale. When stress concentrations develop in Si fibers, primarily at the interfaces and steps on growth TBs, partial dislocations nucleate and glide, leaving SFs behind and relieving the stress. Due to high twin density, this phenomenon is prevalent within the fibers. This phenomenon has been termed as partial dislocation mediated plasticity (PDMP) in Si and is the second attribute of deformation mechanism. Therefore, shearing of Al (CCS) produces high level of strain hardening, and PDMP in Si prevents fiber fracture till a large plastic strain, maintaining mechanical stability of the microstructure.

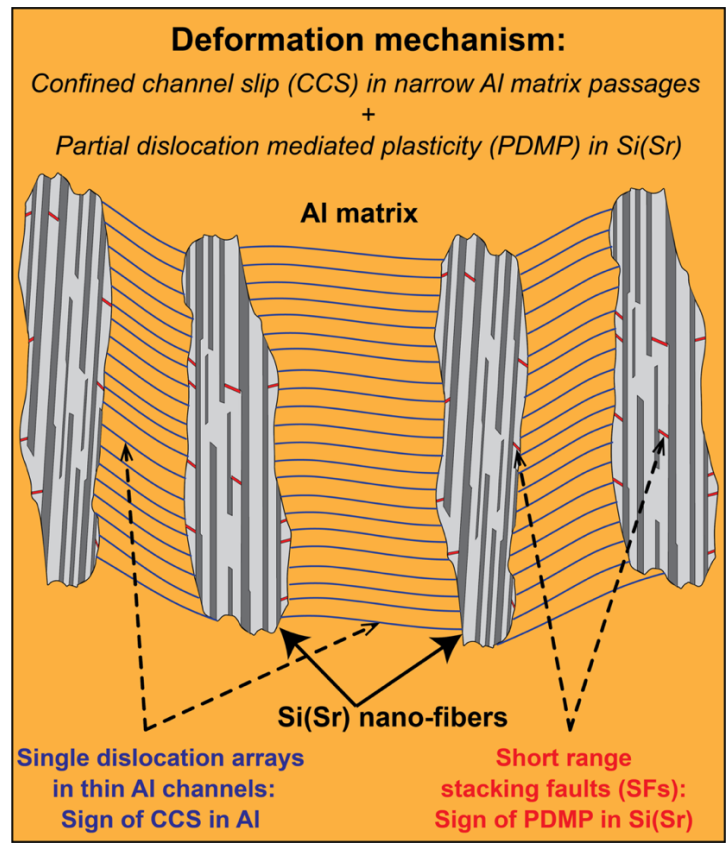

Figure 14: Overall deformation mechanism of the Al-Si(Sr) nano-scale eutectic.

## 5. Summary and Conclusions

The results in this investigation provide direct evidence of enabling plastic co-deformation of Al-Si microstructure at nano-scale, through control of microstructure size, aspect ratio of hard phase, nano-twinning in the hard phase and Sr segregation that may favor partial dislocation mediated plasticity. This work opens up new directions in nanostructure control in laser-processed soft/hard phase microstructures for high flow strength and high plastic deformability.

i) Using a combination of ultra-high cooling rate with 75 μm spot size laser for melting and minor alloying with Sr, unusual microstructures were achieved in Al-20Si alloys that exhibited fully eutectic melt pool microstructure with eutectic colony sizes of several micrometers and Si fibers in the eutectic refined to ≈31 nm, aspect ratio reduced to ≈11 and high twin density in Si fibers with twin thickness/spacing of ≈1-3 nm. Nanoindentation hardness and micropillar compression tests have revealed hardness of up to 2.9 GPa, maximum compressive flow strength of ≈840 MPa with stable plastic flow up to ≈26% plastic strain.

ii) Al-20Si alloy without Sr addition, processed using identical laser parameters, exhibited similar level of refined, aspect ratio and flow strength. However, without Sr alloying, the growth twin density within Si fibers was lower and stable plastic flow reduced to ≈21% strain.

iii) The deformation mechanisms were elucidated by characterizing the deformed microstructures at different length scales. The primary strengthening mechanism involves shearing of nanoscale Al channels in between Si(Sr) fibers resulting in a high density of geometrically necessary dislocations that impart a high strain hardening in Al and favor co-deformation.

iv) Stacking faults were observed in the Si fibers only in the deformed samples, not in the as-laser melted state. The fault density was very high in the Sr-modified deformed microstructure but extremely limited in the laser refined Al-20Si without Sr. Molecular statics reconstruction of the faulted structures confirmed a/6<112> Burgers vector of the partial dislocations. It is hypothesized that limited localized plasticity in the Si fibers is enabled by glide of partial dislocations that are nucleated heterogeneously from sites such as Al/Si interphase boundary or from atomic-level steps on the growth twins. Typically, partial dislocation activity is not observed in monolithic Si crystals, so it is believed that the nanoscale confinement of plasticity in the refined fibrous eutectic develops local stress state that favors partial dislocation nucleation (particularly in Sr-modified eutectics) over cleavage cracking in nanoscale Si fibers that will be investigated in future work using molecular dynamics simulations.

v) Examination of deformation zones underneath nanoindents and compressed micropillars, revealed that Sr-alloyed eutectic completely suppressed cracking of nano-fibers while Al-Si nano-eutectic without Sr exhibits a few cracks in the fibers. The suppression of cracking is presumably due to refinement of Si to diameter of ≈31 nm and reduction of aspect ratio to ≈11 as compared to earlier studies [8] where cracking was frequently observed in Si fibers that were coarser ≈54nm average diameter and higher (≈40) aspect ratio. In the present study, since both alloys, with and without Sr addition, had similar fiber diameter and aspect ratio but the Sr alloyed sample suppressed cracking more effectively, it is postulated that the partial dislocation activity assists in alleviating the local stresses and delaying the onset of cracking in the harder Si phase.

## Acknowledgements

This work is funded by DOE, Office of Science, Office of Basic Energy Sciences with the grant number of DE-SC0016808. Arc melting of the materials were done at the Materials Preparation Center (MPC) at Ames Laboratory, Iowa, USA. Computational work at University of Nebraska - Lincoln was performed in the Nebraska Center for Materials and Nanoscience which is supported by the National Science Foundation under Award ECCS: 1542182 and the Nebraska Research Initiative. Experimental characterization was performed in the Michigan Center for Materials Characterization [(MC)$^2$] at the University of Michigan - Ann Arbor. Authors acknowledge assistance of Dr. Mohsen Taheri Andani in laser scanning experiments and Metin Kayitmazbatir in cooling rate calculations. The authors are also thankful to Dr. Allen Hunter, Dr. Kai Sun, Dr. Tao Ma, and Bobby Kerns for their assistance with experimental methodologies at (MC)$^2$ when needed.

# CRediT Author Statement

**Arkajit Ghosh:** Conceptualization, Methodology, Validation, Formal analysis, Investigation, Data curation, Writing - original draft, Visualization. **Wenqian Wu:** Software, Formal analysis. **Bibhu Prasad Sahu:** Methodology. **Jian Wang:** Writing - review & editing, Project administration. **Amit Misra:** Conceptualization, Writing - review & editing, Supervision, Project administration.

# Declaration of Competing Interest

The authors declare that they have no known competing financial interests or personal relationships that could have appeared to influence the work reported in this paper.

# Data Availability

The data corresponding to this article will be made available upon request. Some part of the data is being used for future works.